\documentclass[aps,prl,twocolumn,superscriptaddress,showpacs,preprintnumbers,amsmath,amssymb,floatfix]{revtex4}

\newcommand{\Aphys}{-1.36}            
\newcommand{\Deltastat}{0.29}         
\newcommand{\Deltasyst}{0.13}         
\newcommand{\Azero}{-2.06}            
\newcommand{\DeltaAzero}{0.14}        
\newcommand{\FakGMs}{0.106}           
\newcommand{\GEsGMs}{0.071}           
\newcommand{\DeltaGEsGMs}{0.036}      
\newcommand{\BeamEnerg}{570.4}        
\newcommand{\Qsquare}{0.108}          
\newcommand{\strangespinofthenucleon}{-0.045}
\newcommand{\strangespinofthenucleonerror}{0.007}
\newcommand{\strangenessdensitylower}{0.22}
\newcommand{\strangenessdensityupper}{0.46}
\newcommand{\strangenesscontributiontonucleonmasslower}{110}
\newcommand{\strangenesscontributiontonucleonmassupper}{220}
\newcommand{\sinthetaweffective}{0.23120(15)}
\newcommand{\radcorasymreduction}{1.3}
\newcommand{\gammafrompizerodilutionfactor}{1}
\newcommand{\weakafourfaklow}{0.106}  
\newcommand{\weakafourfakhigh}{0.225} 
\newcommand{\weakafourresultlow}{0.115}
\newcommand{\weakafourresultlowerror}{0.036}
\newcommand{\weakafourresulthigh}{0.212}
\newcommand{\weakafourresulthigherror}{0.035}
\newcommand{\moeller}{M\o ller}
\newcommand{\latticegmgesvalue}{0.076}
\newcommand{\latticegmgesvalueerror}{0.036}
\newcommand{\samplegmgesvalue}{0.032}
\newcommand{\samplegmgesvalueerror}{0.051}

\newcommand{\cherenkov}{Cerenkov}

\usepackage{graphicx}
\usepackage{dcolumn} 
\usepackage{bm}      
\begin{document}


\title{Evidence for Strange Quark Contributions to the Nucleon's Form
  Factors at $Q^2$ = \Qsquare (GeV/c)$^2$}

\author{F.~E.~Maas}
\email[corresponding author: ]{ maas@kph.uni-mainz.de}
\author{K.~Aulenbacher}
\author{S.~Baunack}
\author{L.~Capozza}
\author{J.~Diefenbach}
\author{B.~Gl\"aser}
\author{T.~Hammel}
\author{D.~von~Harrach}
\author{Y.~Imai}
\author{E.-M.~Kabu\ss}
\author{R.~Kothe}
\author{J.~H.~Lee}
\author{A.~Lorente}
\author{E.~Schilling}
\author{D.~Schwaab}
\author{M.~Sikora}
\author{G.~Stephan}
\author{G.~Weber}
\author{C.~Weinrich}
\affiliation{Institut f\"ur Kernphysik, Johannes Gutenberg-Universit\"at Mainz,
  J.J.~Becherweg~45, D-55099~Mainz, Germany
}
\author{I.~Altarev}
\affiliation{St.~Petersburg Institute of Nuclear Physics, Gatchina, Russia}
\author{J.~Arvieux}
\author{M.~El-Yakoubi}
\author{R.~Frascaria}
\author{R.~Kunne}
\author{M.~Morlet}
\author{S.~Ong}
\author{J.~van~de~Wiele}
\affiliation{Institut de Physique Nucl{\'e}aire (CNRS/IN2P3), Universit{\'e} Paris-Sud, F-91406 - Orsay Cedex, France
}
\author{S.~Kowalski}
\author{B.~Plaster}
\author{R.~Suleiman}
\author{S.~Taylor}
\affiliation{Laboratory for Nuclear Science, Massachussetts Institute of Technology,
Cambridge, MA~02139, USA
}

\date{\today}
\begin{abstract}
We report on a measurement of the parity violating asymmetry in the elastic
scattering of polarized electrons off unpolarized protons with the A4
apparatus at MAMI in Mainz
at a four momentum transfer value of
$Q^2$ = \Qsquare~(GeV/c)$^2$ and at a forward electron scattering angle
of 30$^\circ < \theta_e < 40^\circ$.
The measured asymmetry is $A_{LR}(\vec{e}p)$ = (\Aphys\ $\pm$
\Deltastat$_{stat}$ $\pm$ \Deltasyst$_{syst}$) $\times$ 10$^{-6}$. The expectation from the Standard Model
assuming no strangeness contribution to the vector current is
A$_0$ = (\Azero $\pm$ \DeltaAzero) $\times$ 10$^{-6}$.
We have improved the statistical accuracy by a factor of 3 as compared
to our previous measurements at a higher $Q^2$.
We have extracted the strangeness contribution to the electromagnetic form factors
from our data to be
$G_E^s$ + \FakGMs\ $G_M^s$ = \GEsGMs\ $\pm $ \DeltaGEsGMs\ at $Q^2$~=~\Qsquare~(GeV/c)$^2$.
As in our previous measurement at higher momentum transfer for $G_E^s$ + 0.230 $G_M^s$, we
again find the value for $G_E^s$ + \FakGMs\ $G_M^s$ to be positive, this time at an improved
significance level of 2 $\sigma$.
\end{abstract}

\maketitle

\begin{figure}
\includegraphics[width=0.47\textwidth]{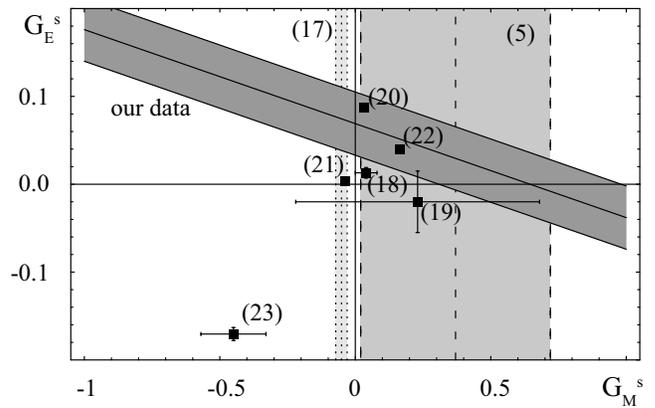}
\caption{\label{fig:GeGm} The solid lines represent our result on
  $G_E^s~+~\FakGMs\ G_M^s$ as extracted from our new data presented here.
  The shaded region represents in all cases the one-$\sigma$-uncertainty
  with statistical and systematic and theory error added in quadrature.
  The dashed lines represent the result on $G_M^s$ from the SAMPLE
  experiment \cite{Spayde04}. The dotted lines represent the result
  of a recent lattice gauge theory calculation for $\mu_s$ \cite{strangeness:leinweber:2004}.
  The black squares represent different model calculations and the numbers denote the references.}
\end{figure}

The understanding of quantum chromo\-dynamics (QCD)
in the nonperturbative regime is crucial for the understanding of the
structure of hadronic matter like protons and neutrons (nucleons).
The successful description of a wide variety
of observables by the concept of effective, heavy ($\approx$ 350~MeV)
constituent quarks, which are not the current quarks of QCD, is still a puzzle.
There are other equivalent descriptions
of hadronic matter at low energy scales in terms of effective fields like
chiral perturbation theory ($\chi$PT) or Skyrme-type soliton models.
The effective fields in these models arise dynamically from
a sea of virtual gluons and quark-antiquark pairs.
In this context the contribution of strange quarks plays a special role
since the nucleon has no net strangeness, and
any contribution to nucleon structure observables
is a pure sea-quark effect.
Due to the higher current mass of the strange quark ($m_s$) as compared
to the up ($m_u$) and down ($m_d$) quark masses with
$m_s \approx 140 \, \mathrm{MeV} \gg  m_u,m_d \approx 5-10 \, \mathrm{MeV}$,
one expects a
suppression of strangeness effects in the creation of
quark-antiquark pairs. On the other hand the strange quark mass is
within the range of the mass scale of QCD ($m_s \approx \Lambda_{QCD}$)
so that the dynamic creation of strange sea quark pairs could still
be substantial with the contribution from c, b, and t-quarks
being negligible.\\
In the past, many hadronic
quantities have been investigated for their strangeness contribution.
For example, the strangeness contribution to the scalar quark density of the vacuum
$\left< 0 | \bar{s}s | 0 \right>$ is sizeable and comparable in magnitude to the
quark condensate of the u and d flavors $\left<0 | \bar{q}q | 0\right>$, namely
$\left<0 | \bar{s}s | 0\right>$ =
(0.8 $\pm$ 0.1) $\left<0 | \bar{q}q | 0\right>$ \cite{strangevacuum:ioffe:2003}.
The scalar strangeness density $\left<N | \bar{s}s | N\right>$ of the nucleon gives a contribution
to the mass of the nucleon. It has been discussed in the context of the $\Sigma$ commutator,
which can be related to the $\pi$N scattering amplitude. 
Recent evaluations of the $\pi$N sigma term from
new $\pi$N scattering data give values
for $\Sigma_{\pi N}$ of (64 $\pm$ 8) MeV up to (79 $\pm$ 7) MeV
resulting in a strangeness ratio $y=2\left<N|\bar{s}s|N\right>/\left<N|\bar{u}u+\bar{d}d|N\right>$
in the range between $\strangenessdensitylower$ and $\strangenessdensityupper$. This corresponds to a
contribution to the nucleon mass
from the scalar density $m_s \left<N|\bar{s}s|N\right>$ of
\strangenesscontributiontonucleonmasslower\ MeV to
\strangenesscontributiontonucleonmassupper\ MeV \cite{strangenesssigma:pavan:2002}.
Information on the axial charge $\left<N | \bar{s}\gamma_\mu \gamma^5 s | N\right>$ and on the
strangeness contribution to the spin of the nucleon
comes from the interpretation of deep inelastic scattering data
and suggests a sizeable contribution of the strange quarks of
$\Delta s$($Q^2$ = 1 (GeV/c)$^2$) = \strangespinofthenucleon\ $\pm$ \strangespinofthenucleonerror\ to the nucleon
spin from a next-to-leading order perturbative QCD analysis of the available world data set
including higher twist effects \cite{strangespin:leader:2003}.\\
Parity violating (PV) electron scattering off nucleons
provides experimental access to the strange quark vector current in the
nucleon $\left<N | \bar{s} \gamma_\mu s | N\right>$, which is parameterized
in the electromagnetic form factors of proton and neutron,
$G_E^s$ and $G_M^s$ \cite{pvstrangeness:kaplan:88}.
Recently three collaborations have published experimental
results: 1. The SAMPLE collaboration at MIT-Bates \cite{Spayde04, Ito04}
at a four momentum transfer of $Q^2$~=~0.1~(GeV/c)$^2$ and
$Q^2$~=~0.04~(GeV/c)$^2$ at backward angles, where they are
sensitive to $G_M^s$ and the axial form factor
$\tilde{G}_A^p$, 2. the HAPPEX collaboration at TJNAF \cite{Happex04}
with scattering at a momentum transfer of $Q^2$ = 0.48 (GeV/c)$^2$
at forward angles of 12$^\circ$ where they are sensitive to
$G_E^s$~+~0.39~$G_M^s$, and 3. our collaboration A4 at MAMI
at a momentum transfer of $Q^2$~=~0.23~(GeV/c)$^2$
with scattering at forward angles sensitive to
$G_E^s$~+~0.25~$G_M^s$. A direct separation of the electric ($G_E^s$) and magnetic
($G_M^s$) contributions at forward angles has been impossible so far,
since the measurements have been taken at different $Q^2$-values.
We present here a new measurement of the PV asymmetry in
the scattering of polarized electrons off unpolarized
protons with the A4 experiment at forward angles of
$30^\circ < \theta_e < 40^\circ$ and at the same four momentum
transfer of $Q^2$~=~\Qsquare~(GeV/c)$^2$ where the SAMPLE collaboration
has measured at backward kinematics so that for the
first time a separation of $G_E^s$ and $G_M^s$ solely from
experimental data will be possible.\\
The interference between weak ($Z$) and electromagnetic
($\gamma$) amplitudes
leads to a PV asymmetry $A_{LR}(\vec{e}p)=(\sigma_R-\sigma_L)/(\sigma_R+\sigma_L)$
in the elastic scattering cross section of
right- and left-handed electrons ($\sigma_R$ and $\sigma_L$ respectively).
The asymmetry can be written as
$A_{LR}(\vec{e}p)$ = c$_1$ $\tilde{G}_E^p$ + c$_2$ $\tilde{G}_M^p$ + c$_3$ $\tilde{G}_A^p$ and
is given in the framework of the Standard Model \cite{pvsummary:musolf:94}.
The weak vector form factors $\tilde{G}^p_{E,M}$
of the proton can be expressed in terms of the known electromagnetic nucleon
form factors $G_{E,M}^{p,n}$ and the unknown strange form factors
$G^s_{E,M}$ using isospin symmetry and the universality of the quarks in weak and
electromagnetic interactions. It can be
expressed as a sum of three terms, $A_{LR}(\vec{e}p) = A_V + A_s + A_A $.
\begin{eqnarray}
A_V=-a \rho'_{eq}\{(1-4\hat{\kappa}'_{eq}\hat{s}^2_Z)
     -\frac{\epsilon G_E^p G_E^n + \tau G_M^p G_M^n}{\epsilon (G_E^p)^2 + \tau
     (G_M^p)^2}\}, \label{eq:Av} \\
A_s=a{\rho'_{eq}\frac{\epsilon G_E^p G_E^s + \tau G_M^p G_M^s}{\epsilon
(G_E^p)^2 + \tau (G_M^p)^2}}, \label{eq:As}\\
A_A=a{\frac{(1-4\hat{s}^2_Z)\sqrt{1-\epsilon^2}{\sqrt{\tau(1+\tau)}G_M^p\tilde{G}_A^p}}{\epsilon
(G_E^p)^2 + \tau (G_M^p)^2}}. \label{eq:Aa}
\end{eqnarray}
$A_{V}$ represents the vector coupling at the proton vertex where the possible
strangeness contribution has been taken out and has been put into $A_s$,
a term arising only from a contribution of strangeness to the
electromagnetic vector form factors.
The term $A_A$ represents the contribution from the axial coupling at the proton
vertex due to the neutral current weak axial form factor $\tilde{G}_A^p$.
The quantity $a$ represents $(G_{\mu} Q^2) /(4 \pi \alpha \sqrt{2})$.
$G_{\mu}$ is the Fermi coupling constant as derived from muon decay.
$\alpha$ is the fine structure constant,
$Q^2$ the negative square of the four momentum transfer,
$\tau = Q^2/(4 M_p^2)$ with $M_p$ the proton mass, and
$\epsilon = [1 + 2 (1 + \tau)\tan^2(\theta_e/2)]^{-1}$
with $\theta_e$ the laboratory scattering angle of the electron.
The electromagnetic form factors $G_{E,M}^{p,n}$ are taken from a recent
parametrization (version 1, page 5) by Friedrich and Walcher \cite{emformfactor:friedrich:03},
where we assign an experimental error of $3\,\%$ to $G_M^{p}$ and $G_E^p$,
$5\,\%$ to $G_M^n$, and
$10\,\%$ to $G_E^n$.
%
%
Electro-weak radiative corrections are included in the factors $\rho_{eq}'$
and $\hat{\kappa}_{eq}'$ which have been evaluated in the $\overline{MS}$
renormalization scheme according to \cite{ewcorrections:marciano:84}.
We use a value for $\hat{s}^2_Z = \sin^2 \hat{\theta}_W(M_Z)_{\overline{MS}}$ of
\sinthetaweffective\   \cite{pdg:eidelmann:04}.
\begin{figure}
\includegraphics[width=0.47\textwidth]{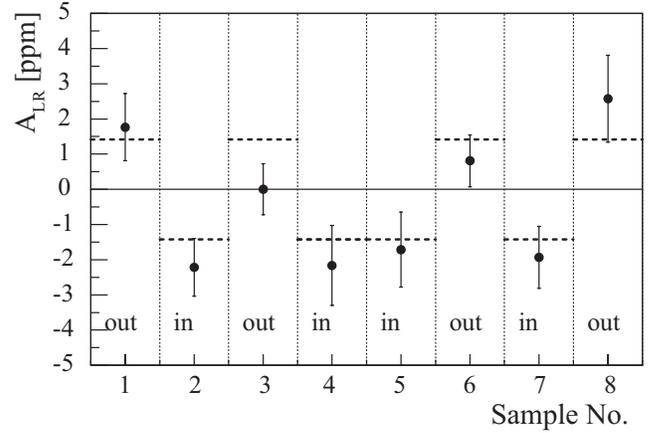}
\caption{\label{fig:SamplePlot}Extracted physics asymmetries for $Q^2$ = \Qsquare (GeV/c)$^2$ with respect
  to the state of the $\lambda$/2 wave plate. The flip of sign can be observed
  well. The dashed line represents the global fit to all data with $\lambda/2$-wave
  plate sign flip taken into account.}
\end{figure}
The electro-weak radiative corrections to $A_A$ \cite{axialformfactor:zhu:00}
are included in $\tilde{G}_A^p$.
Electromagnetic internal and external radiative corrections to the asymmetry
and the effect of energy loss due to ionization in the target have also been
calculated. They reduce the expected asymmetry for our kinematics by about
$\radcorasymreduction\,\%$.
We average $A_0 = A_{V} + A_{A}$
over the acceptance of the detector and the target length. We obtain
an expected value for the asymmetry at the averaged $Q^2$ without
strangeness contribution to the vector form factors of
$A_0 = (\Azero \pm \DeltaAzero) \times 10^{-6}$.
The three largest contributions to the uncertainty of $A_0$ come from the uncertainty
in the axial form factor
$\tilde{G}_A^p$ ($0.08 \times 10^{-6}$), the electric form factor of the proton $G_E^p$ ($0.07 \times 10^{-6}$),
and the magnetic form factor of the neutron $G_M^n$ ($0.07 \times 10^{-6}$).\\
The measurement was carried out at the MAMI accelerator facility using the setup of the A4 experiment
\cite{MAMI:euteneuer:94,a4calori:maas:02,a4experi:maas:03,a4experiment:maas:2004}.
We used an electron beam with an energy of \BeamEnerg\ MeV and an intensity of $20~\mu$A.
An averaged electron beam polarization $P_e$ of 80 \% was obtained
using a strained layer GaAs
crystal. The helicity of the electrons is changed randomly between two complementary
patterns on a 20~ms basis.
The beam polarization was measured weekly by a \moeller\
polarimeter with a precision of better than 2~\%. Together with the interpolation between the weekly measurements
the uncertainty in the knowledge of the polarization was 4~\%. Several monitor and stabilization systems
were installed along the accelerator to minimize and measure
helicity correlated beam fluctuations such as differences in
position, angle, and energy for the two helicity states, which introduce false asymmetries.
The electrons were scattered off a 10~cm long liquid hydrogen target
corresponding to a luminosity of $L~=~0.5~\times~10^{38}~cm^{-2}s^{-1}$,
monitored by eight water-\cherenkov\ detectors mounted at small
scattering angles of $4.4^{\circ} < \theta_e < 10^{\circ}$ covering the full azimuthal
range of $\phi_e$.
From the helicity correlated luminosity and beam current measurement we obtained the target
density $\rho_{R,L}~=~L_{R,L}/I_{R,L}$ for the two helicity states R and L from the ratio
of luminosity $L_{R,L}$ and beam current $I_ {R,L}$.\\
\begin{table}
  \caption{\label{tab:Korrekturen570MeV}Applied corrections to the measured asymmetry and their
  contribution to the systematic error at $Q^2$ = \Qsquare (GeV/c)$^2$ in units of parts per million (ppm).}

  \begin{ruledtabular}
    \begin{tabular}{l r r}
                                       & Correction (ppm)& Error (ppm)\\ \hline
      Target density, luminosity       & -0.32           & 0.01\\
      Target density, current          &  0.00           & 0.02\\
      Nonlinearity of LuMo             &  0.05           & 0.00\\
      Dead time correction             &  0.03           & 0.02\\
      Current asymmetry $A_I$          & -0.33           & 0.04\\
      Energy difference $\Delta E$     &  0.03           & 0.03\\
      Position differences $\Delta x$,$\Delta y$&  0.01  & 0.09\\
      Angle differences $\Delta x'$, $\Delta y'$&  0.01  & 0.09\\
      Al windows (H$_2$ target)        &  0.06           & 0.03\\
      Dilution from $\pi^0$ decay      &  0.00           & 0.02\\
      $P_e$ measurement                & -0.34           & 0.03\\
      $P_e$ interpolation              &  0.00           & 0.05\\ \hline
      Total systematic error           &                 & 0.13\\
    \end{tabular}
  \end{ruledtabular}
\end{table}
The scattered particles were detected in a total absorbing calorimeter that
consisted of 1022 individual lead fluoride (PbF$_2$) crystals placed in 7 rings
and 146 rows. It covered scattering angles from $30^{\circ} < \theta_e < 40^{\circ}$
and the full azimuthal $\phi_e$-range, yielding a solid angle of $\Delta\Omega~=~0.62$~sr.
We achieved an energy resolution of $3.9\%/\sqrt{E}$ at the elastic peak.
This allows a clean separation of elastic scattered
electrons from inelastic events like $\pi$-electroproduction and $\Delta(1232)$-excitation,
which have a PV asymmetry originating from different physics.
Using Monte Carlo simulations we estimated the possible background contribution
from the production of
$\pi^0$s, which subsequently decay into two photons,
to be much less than
$\gammafrompizerodilutionfactor\,\%$. It is neglected here.
The largest background comes from quasi-elastic scattering off the thin aluminum entrance
and exit windows of the target cell (see Table \ref{tab:Korrekturen570MeV}).\\
The number of elastic events for positive (R) and negative (L) helicity $N_{R,L}$ was
determined from the energy histograms and from summing
up over all 730 channels of the inner five rings of the calorimeter. For each
5 minute run we calculated a raw asymmetry using the elastic counts normalized to the
target density $A_{raw}=(N_R/\rho_R - N_L/\rho_L)/(N_R/\rho_R + N_L/\rho_L)$.
Corrections due to false asymmetries arising from helicity correlated changes of beam parameters
were applied on a run by run basis, using
the method of multilinear regression. We corrected for the measured polarization $P_e$
of the electron beam.
The analysis is based on a total of $4.8 \times 10^6$ histograms
corresponding to $2 \times 10^{13}$ elastic events.
The largest correction arises from the interpolation between the polarization measurements
and the corrections
for the beam current asymmetry. Table \ref{tab:Korrekturen570MeV} gives an overview of the
corrections and their contributions to the systematic error.
About half of the data samples were taken with a $\lambda/2$-wave plate inserted in the
\begin{figure}
\includegraphics[width=0.47\textwidth]{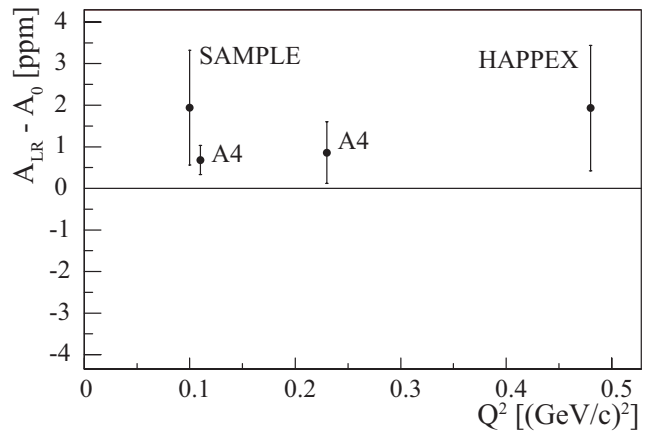}
\caption{\label{fig:DiffA0}Difference between the measured parity violating asymmetry
  in electron proton scattering
  $A_{LR}(\vec{e}p)$ and the asymmetry $A_0$ without vector
  strangeness contribution
  from the Standard Model for the SAMPLE experiment at backward angles,
  and the HAPPEX and the two A4 results at forward angles.
  We have taken the published asymmetry values from the SAMPLE and HAPPEX
  results and recalculated $A_0$ according to Equations \ref{eq:Av} and \ref{eq:Aa},
  taking for SAMPLE the mean values of the kinematical variables.
  The new experimental result at $Q^2$=\Qsquare~(GeV/c)$^2$ presented here
  is the most accurate measurement.}
\end{figure}
laser optics of the electron source which reverses the beam helicity from R to
L and vice versa without any other change to the data acquisition system or the
analysis. This serves as a sensitive test for the understanding of the systematic
corrections since the extracted physics asymmetry $A_{LR}(\vec{e}p)$ must change sign.
Altogether eight so called data samples have been
taken, four with the $\lambda/2$-wave plate in. Figure \ref{fig:SamplePlot} shows the
extracted physics asymmetries for these data samples. The change in sign when
the $\lambda/2$-wave plate was inserted can easily be observed.
From all of the data we extracted a value of $A_{LR}(\vec{e}p)=(\Aphys \pm
\Deltastat_{stat} \pm \Deltasyst_{syst})$ ppm.\\
A combination of the weak form factors
of the proton $\tilde{G}_E^p$ and $\tilde{G}_M^p$ can be extracted from the
measured asymmetry.
For the extraction we take $\tilde{G}_A^p$ from \cite{axialformfactor:zhu:00}.
At our average momentum transfer of $Q^2$ = \Qsquare\ (GeV/c)$^2$ we extract a value of
$\tilde{G}_E^p$ + \weakafourfaklow\ $\tilde{G}_M^p$ = \weakafourresultlow\ $\pm$ \weakafourresultlowerror.
We compute the result for our previous measurement at
the higher $Q^2$ of 0.230 (GeV/c)$^2$ and yield
$\tilde{G}_E^p$ + \weakafourfakhigh\ $\tilde{G}_M^p$ = \weakafourresulthigh\ $\pm$ \weakafourresulthigherror,
which has been omitted in \cite{a4experiment:maas:2004}.\\
The statistical and systematic error of the measured asymmetry and the error
in the theoretical prediction of $A_0$ are added in quadrature.
The difference between the measured value of $A_{LR}(\vec{e}p)$
and the Standard Model prediction without vector strangeness contribution
is directly proportional to the strangeness contribution to the
vector form factors of the nucleon.
The previously reported results of SAMPLE, HAPPEX and A4
at a different momentum transfer give a measured value for $A_{LR}(\vec{e}p)$ smaller than $A_0$.
For our new data presented here the difference of $A_{LR}(\vec{e}p)$ - $A_0$
has a two $\sigma$ deviation from zero.
Each single measurement does not give a strangeness signal larger that two $\sigma$,
but the combination of the three forward angle measurements seems to show
evidence for the observation of strangeness effects.
We have taken the published asymmetry values from the SAMPLE and HAPPEX
results and recalculated $A_0$ according to Equations \ref{eq:Av} and \ref{eq:Aa},
taking for SAMPLE the mean values of the kinematical variables.
Figure \ref{fig:DiffA0} shows the present situation of all the published
asymmetry measurements in elastic electron proton scattering
using our parameter set for the calculation of $A_0$.\\
From the difference between the measured $A_{LR}(\vec{e}p)$ and
the theoretical prediction in the framework of the Standard Model, $A_0$, we
extract a linear combination of the strange electric and magnetic form factors
of $G_E^s$ + \FakGMs\ $G_M^s$ = \GEsGMs\ $\pm$ \DeltaGEsGMs\ at $Q^2$ = \Qsquare\ (GeV/c)$^2$.
In Figure \ref{fig:GeGm}, the solid lines show the possible values of $G_E^s$ against $G_M^s$, the
gray shaded area represents the 65 \% confidence limit error band.
Our result is about two standard deviations away from
$G_E^s+\FakGMs\ G_M^s = 0$. The gray shaded area with the dashed lines is
the result from the SAMPLE experiment on $G_M^s$ at the same $Q^2$ as measured
at backward scattering angles.
If we take the published SAMPLE result on $G_M^s$, we get a value for
$G_E^s$ = \samplegmgesvalue\ $\pm$ \samplegmgesvalueerror.
A consistent combination of our result on $G_E^s$ + \FakGMs\ $G_M^s$ presented here and
the SAMPLE result on $G_M^s$ would require a
reanalysis of averaging $A_0$ over the detector acceptance with
a common set of electromagnetic and axial form factors.
Work is in progress on a common A4/SAMPLE analysis.
A recent very accurate determination of the strangeness contribution
to the magnetic moment of the proton $\mu_s = G_M^s(Q^2$ = 0 (GeV/c)$^2$)
from lattice gauge theory \cite{strangeness:leinweber:2004}
is indicated by the dotted lines. It would yield a larger value
of $G_E^s$ = \latticegmgesvalue\ $\pm$ \latticegmgesvalueerror\
if the $Q^2$ dependence of $G_M^s$ from 0 to \Qsquare~(GeV/c)$^2$ were neglected.
The theoretical expectations from another quenched lattice gauge theory calculation
\cite{strangeness:lewis:2003}, from SU(3) chiral perturbation theory
\cite{strangeness:meissner:1999}, from a chiral soliton model
\cite{strangeness:goeke:2002}, from a quark model
\cite{strangeness:faessler:2002}, from a Skyrme-type soliton model
\cite{strangeness:weigel:95} and from an updated vector meson dominance model
fit to the form factors \cite{strangeness:hammer:1999} are depicted
into Figure \ref{fig:GeGm}. \\
We are preparing a series of measurements of the parity violating
asymmetry in the scattering of longitudinally polarized
electrons off unpolarized protons and deuterons at
backward scattering angles of $140^\circ < \theta_e < 150^\circ$
with the A4 apparatus in order to separate the electric ($G_E^s$)
and magnetic ($G_M^s$) strangeness contribution to the
electromagnetic form factors of the nucleon.\\
%
%
%
%
This work is supported by the DFG
under SFB 201, SPP 1034, by the IN2P3 of CNRS
and in part by the DOE.
We are indebted to K.H. Kaiser and the MAMI crew and we
thank the A1 Collaboration for the \moeller\ polarimeter
measurements. We would like to thank S.~Scherer
and H.~Spiessberger for useful discussions.

\bibliography{570long}

\end{document}